\documentstyle[prb,aps,multicol,epsf]{revtex}
\input psfig
\def\be{\begin{equation}}
\def\ee{\end{equation}}

\def\De{\Delta}
\def\sqr#1#2{{\vcenter{\vbox{\hrule height .#2pt
      \hbox{\vrule width .#2pt height#1pt \kern#1pt
      \vrule width.#2pt}
      \hrule height.#2pt}}}}
\def\square{\mathchoice\sqr34\sqr34\sqr{2.1}3\sqr{1.5}3}

\begin{document}
\bibliographystyle{simpl1}

\title{The Upper Critical Field in Disordered Two-Dimensional
Superconductors}

\author{Robert A Smith$^{1,2}$, Beccy S Handy$^{1}$ and Vinay Ambegaokar$^{2}$}

\address{$^{1}$School of Physics and Astronomy, University of
Birmingham, Edgbaston, Birmingham B15 2TT, ENGLAND}

\address{$^{2}$Laboratory of Atomic and Solid State Physics,
Cornell University, Ithaca, New York 14853, USA}
\maketitle
\bigskip

\begin{abstract}
We present calculations of the upper critical field $H_{c2}(T)$
in superconducting films as a function of increasing disorder
(as measured by the normal state resistance per square $R_{\square}$).
In contradiction to previous work, we find that there is no anomalous
low temperature positive curvature in $H_{c2}$ vs $T$ as disorder is
increased. We show that the previous prediction of this effect is due
to an unjustified analytical approximation of sums occuring in the
perturbative calculation. Our treatment includes both a careful
analysis of first-order perturbation theory, and a non-perturbative
resummation technique. No anomalous curvature is found in either
case. We present our results in graphical form.
\end{abstract}

\section{Introduction}

Increasing disorder is known to suppress superconductivity in
low-dimensional systems such as thin films and narrow 
wires\cite{MF,TK,Fink87,SRW}. This
occurs because the disorder causes electrons to move diffusively
rather than ballistically, making them less efficient at screening
the Coulomb repulsion between electrons. The increased Coulomb repulsion
decreases both the electron-electron attraction needed for superconductivity,
and the density of states of electrons available for pairing at the Fermi
surface\cite{Fink94}. 
Typical types of experimental data are: (i) $T_c(R_{\square})$, the
transition temperature as a function of normal state resistance
per square\cite{RLCM,GB,HLG,LK,XHD}; 
(ii) $\De_0(R_{\square})$, the order parameter at zero temperature,
as a function of normal state resistance per square\cite{VDG,Vall94};
(iii) $H_{c2}(T,R_{\square})\equiv T_c(R_{\square},H)$, the upper critical
field as a function of temperature and normal state resistance per
square\cite{GB,HP,OKOK};
(iv) $T_c(R_{\square},1/\tau_s)$, transition temperature as a function of
resistance per square and spin-flip scattering rate in films with magnetic
impurities.\cite{CV} It is found experimentally that $T_c(R_{\square})$ curves
from a wide variety of
materials fit a universal curve with a single fitting parameter, whilst the
few experimental measurements of $\Delta_0(R_{\square})$, seem to have
$\Delta_0(R_{\square})/2k_BT_c(R_{\square})$ roughly 
constant\cite{VDG,Vall94}. This fitting to a single
curve, whilst pleasing in showing that the basic ingredients of our theories
are correct, does not allow detailed analysis of the theory. Data of types
(iii) and (iv) are more promising because there is an additional parameter
to vary -- the magnetic field in (iii), and spin-flip scattering rate in (iv).
To the best of our knowledge, only one experiment of type (iv) has been
performed\cite{CV}, and we discuss it elsewhere.\cite{AS99} 
Several experiments of type (iii)
have been performed\cite{GB,HP,OKOK}; some seem to show a positive curvature in 
$H_{c2}(T)$ at low temperature
as disorder is increased. Moreover this effect is predicted by 
theory\cite{MEF,GD90}, and
this seems to be another confirmation of the basic theoretical model. 
However, we need to be careful: 
positive curvature in $H_{c2}(T)$ is a ubiquitous
feature
of exotic superconductors\cite{Bran98}, 
and occurs in many systems where localization is
not believed to be the cause. Indeed any pair-breaking mechanism that varies
as a function of magnetic field can lead to such anomalous behaviour in
$H_{c2}$.
This means that it is often difficult to distinguish between the various
mechanisms that might be present. It is therefore particularly important to
be sure of our theory, and in this light we re-examine the predictions of
localization theory.
\par
One of the main problems of the localization theory is that even
first-order perturbation theory results are hard to obtain correctly.
The first-order results are capable of explaining experimental data
in the weak disorder regime, but for stronger disorder it is clear that
we need something else. As an example consider the prediction for $T_c$
suppression\cite{MF,TK},
\begin{equation}
\label{mftc}
\ln{\left({T_c\over T_{c0}}\right)}=-{1\over 3}{R_{\square}\over R_0}
\ln^3{\left({1\over 2\pi T_{c0}\tau}\right)}
\end{equation}
where $T_{c0}$ is transition temperature for clean system,
$R_0=2\pi h/e^2\approx 162k\Omega$, and $\tau$ is the elastic scattering time.
This yields an exponential curve for $T_c(R_{\square})$, which behaves
like a straight line for small $R_{\square}$. It is clear
that $T_c$ deduced from this equation can never go to zero for finite
$R_{\square}$, as happens in experiment. A very simple ad hoc way of
going beyond simple perturbation theory is to replace $T_{c0}$ on
the right hand side by $T_c$, pleading perhaps to self-consistency.
If we define $x=\ln{(T_{c0}/T_c)}$, $\beta=\ln{(1/2\pi T_{c0}\tau)}$,
and $t=R_{\square}/R_0$, the new equation has the cubic form
\begin{equation}
\label{cubic}
x={t\over 3}(\beta+x)^3
\end{equation}
and can easily be solved. However a new problem emerges because there
are two positive roots for every value of $R_{\square}$. At first we
can take the larger of the roots, on physical principles, because it is this
root which tends to $T_c/T_{c0}=1$ at $R_{\square}=0$. However we
eventually come to a
re-entrance point beyond which no solutions exist. It is clear that this
re-entrance is unphysical, an artefact of our ad hoc extension of
perturbation theory. In the case of $T_c(R_{\square})$ the story has a
happy ending in that perturbation theory can be correctly extended by
a renormalization group (RG) treatment based on Finkelstein's interacting
non-linear sigma model\cite{Fink87,Fink83}. This leads to the result
\begin{equation}
\label{rgtc}
\ln{{\left(T_{c}\over T_{c0}\right)}}=
{1\over |\gamma|}-{1\over 2\sqrt{t}}
\ln{\left({1+\sqrt{t}/|\gamma|\over 1-\sqrt{t}/|\gamma|}\right)},
\end{equation}
where $\gamma=-1/\ln{(T_{c0}/1.13\tau)}$. This equation reduces to the
first-order result for small $t$ and now $T_c$ goes smoothly to zero
at $t=\gamma^2$. The three curves are plotted for comparison in Fig. (1).
\begin{figure}
\centerline{\psfig{figure=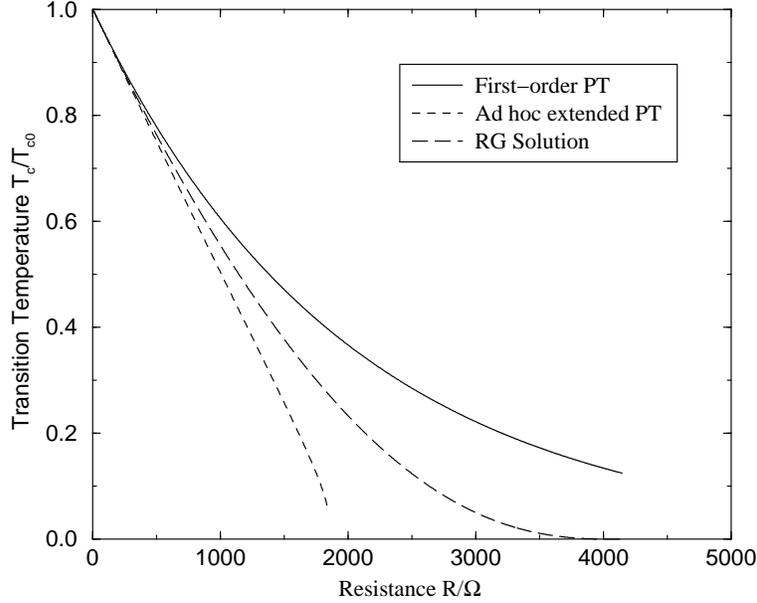,width=10cm}}
\caption{Theoretical predictions for transition temperature as a function
of resistance per square. The solid line is the first-order prediction;
the dashed line is the ``self-consistent'' first-order prediction, which 
suffers from re-entrance; the long-dashed line is the result obtained from
either renormalization group or ladder re-summation approaches.}
\end{figure}
\par
The reason for discussing $T_c(R_{\square})$ in detail above is that the
same problem occurs for $H_{c2}(T,R_{\square})\equiv T_c(R_{\square},H)$.
The standard theory in this case\cite{MEF}, due to 
Maekawa, Ebisawa and Fukuyama (MEF)
is the equivalent of the ad hoc
extension discussed above, and has the form
\begin{eqnarray}
\label{mef}
\ln{\left({T_{c}\over T_{c0}}\right)}&=&\displaystyle
\psi\left({1\over 2}\right)-\psi\left({1\over 2}+{1\over 2\pi T_c\tau_H}
\left[1-2t\ln{\left({1\over 2\pi T_c\tau}\right)}\right]\right)-R_{HF}-R_V
\nonumber\\
R_{HF}&=&\displaystyle-{1\over 2}t\ln^2{\left({1\over 2\pi T_c\tau}\right)}
-t\ln{\left({1\over 2\pi T_c\tau}\right)}
\left[\psi\left({1\over 2}\right)
-\psi\left({1\over 2}+{1\over 2\pi T_c\tau_H}\right)\right]\nonumber\\
R_V&=&\displaystyle-{1\over 3}t\ln^3{\left({1\over 2\pi T_c\tau}\right)}
-t\ln^2{\left({1\over 2\pi T_c\tau}\right)}
\left[\psi\left({1\over 2}\right)
-\psi\left({1\over 2}+{1\over 2\pi T_c\tau_H}\right)\right],
\end{eqnarray}
where
$1/\tau_H=DeH$. This equation suffers similar re-entrance problems at
finite $1/\tau_H$ to those found at $1/\tau_H=0$, when it is just the $T_c$
equation. Indeed the $H_{c2}(T)$ curves can only be plotted down to
the value of $T$ at which re-entrance occurs, and at this point the
curves appear to have infinite slope. This leads to us asking the
question of whether
the positive curvature in $H_{c2}(T)$ is also an artefact of the ad hoc
approximation used. What we need to answer this question is the finite
magnetic field analogue of the RG result discussed above. However the
RG is very difficult, and the answer is not forthcoming from this source.
Fortunately Oreg and Finkelstein\cite{OF} have recently shown that the RG result
can be obtained from a diagrammatic resummation technique. This method has
the great advantage of being analytically tractable and easy to use.
In this paper
we extend this approach to finite magnetic field to see whether we really
do expect positive curvature in $H_{c2}(T)$. Indeed this paper is intended
as somewhat of a showcase for this resummation method, to demonstrate the
ease of its extension to a wide variety of problems.
\par
Oreg and Finkel'stein\cite{OF} take the Coulomb interaction to be featureless,
and we shall follow them. We should therefore explain why it is legitimate
to use a featureless
interaction rather than the correct screened Coulomb interaction. The
screened Coulomb interaction has a singularity at low momentum, and one
might naively think that this would lead to a strong enhancement of the
suppression of transition temperature.
However a cancellation occurs between diagrams 1--4 and diagram 5 of Fig. 2,
which effectively removes the low momentum singularity\cite{Fink87,SRW}. 
The Coulomb interaction
is then effectively featureless, all diagrammatic sums being dominated by
large frequency and momentum. It turns out that one gets the same result
from doing the perturbation theory correctly with the screened Coulomb
interaction as from using a constant interaction of strength $g=N(0)V=1/2$
and keeping only diagrams
1--4. This is what we shall do in this paper. We also note that the
re-summation method of Oreg and Finkelstein\cite{OF} keeps only 
diagrams 3 and 4,
which is again legitimate as their contribution is greater than that from
diagrams 1 and 2. However it is not difficult to include these terms in the
re-summation, as we will show in section
\uppercase\expandafter{\romannumeral3}. The cancellation of low momentum
singularities is expected to be a general feature, enforced by gauge
invariance, so we are free to ignore them and use a featureless interaction
as long as we include the diagrams that give the dominant contribution at
large frequency and momentum.
\begin{figure}[t]
\centerline{\psfig{figure=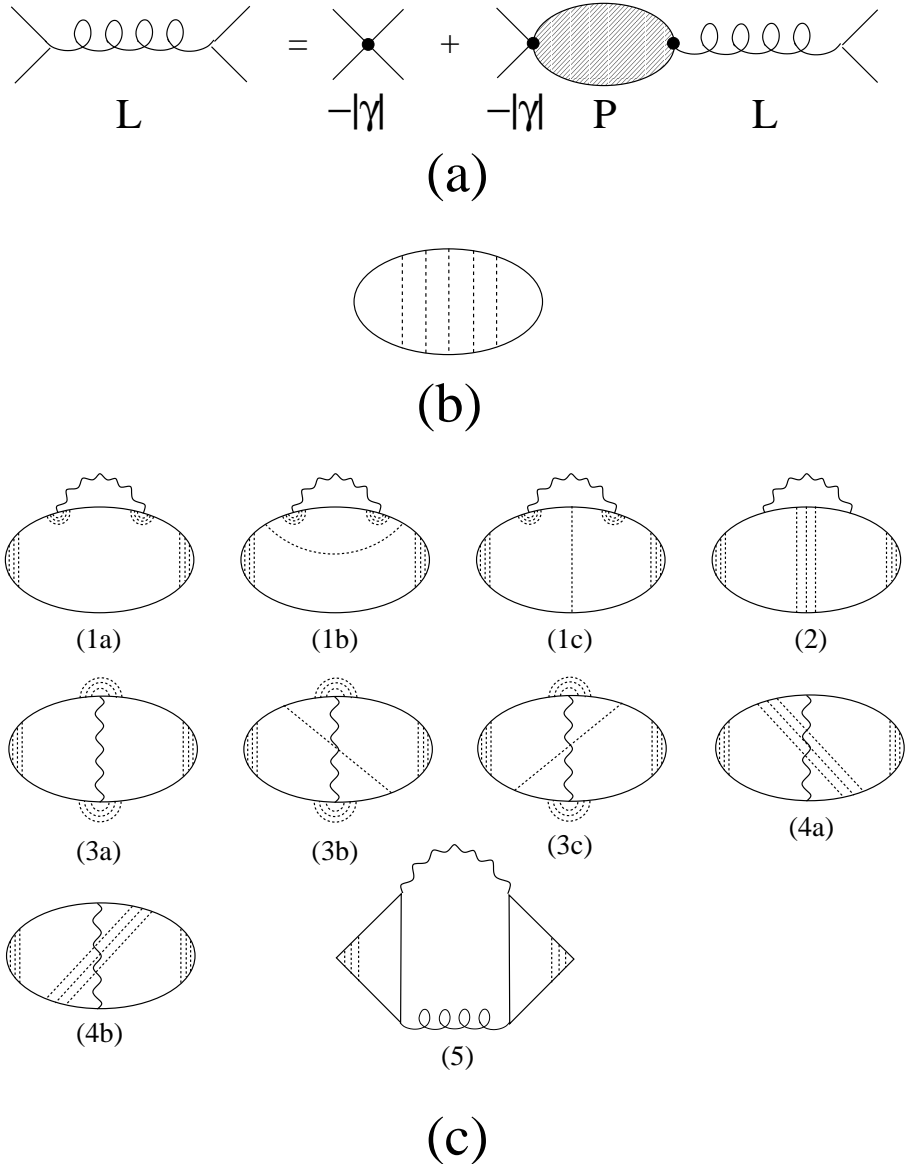,width=12cm}}
\medskip
\caption{The diagrams needed to calculate the first-order perturbative
correction to the pair-propagator $L(q,0)$. (a) Definition of pair propagator
$L(q,0)$ in terms of the pair polarization bubble $P(q,0)$ and the 
4-point BCS
interaction $-|\gamma|$. (b) Zeroth-order (mean-field) polarization bubble
in a dirty superconductor. The solid lines are electron Green functions;
the dashed lines are impurity lines. (c) First-order perturbative correction
to pair polarization bubble. The wiggly line is the screened Coulomb
interaction.}
\end{figure}
\par
The outline of the rest of the paper is as follows. In section
\uppercase\expandafter{\romannumeral2} we review the first-order
perturbation calculation for $H_{c2}(T)$ assuming a featureless
Coulomb interaction. We derive the analytical MEF formula by making
an asymptotic approximation to the Matsubara frequency sums which arise
in perturbation theory. We make plots of
$H_{c2}(T)$ both by solving the implicit MEF equation, and its equivalent
where the Matsubara sums are performed exactly.
Surprisingly we find no positive curvature in the latter case. In
section \uppercase\expandafter{\romannumeral3} we derive $H_{c2}(T)$
using the resummation technique both in the form used by Oreg and
Finkel'stein, and also in an extended form which includes self-energy
diagrams missed in their formalism. Again we find no positive curvature
in $H_{c2}(T)$. In section \uppercase\expandafter{\romannumeral4} we
discuss the experimental situation and draw conclusions.

\section{Review of First-Order Perturbation Theory}

In this section we will carefully review the first-order perturbation
theory calculation\cite{MEF} of $H_{c2}(T)$. 
We do this in detail because we will
show that accurately performing sums rather than making an analytical
approximation to them removes the upward curvature in $H_{c2}(T)$. We
identify $T_c$ as the temperature at which the magnetic field dependent
pair propagator diverges. To find the latter we calculate the correction
to the momentum-dependent pair
propagator, $L(q,0)$, and then make the usual substitution
$Dq^2\rightarrow 2/\tau_H$, where $1/\tau_H=DeH$. The zeroth-order (mean
field) pair propagator is given by
\begin{equation}
\label{l0def}
L_0(q,0)^{-1}=N(0)\left[\ln{\left({T\over T_{c0}}\right)}
+\psi\left({1\over 2}+{Dq^2\over 4\pi T}\right)
-\psi\left({1\over 2}\right)\right]
\end{equation}
so that upon substitution $Dq^2\rightarrow 2/\tau_H$ we get the usual
Abrikosov-Gorkov result\cite{AG} for the pair-breaking effect of the magnetic
field
\begin{equation}
\label{agtc}
\ln{\left({T_c\over T_{c0}}\right)}=
\psi\left({1\over 2}\right)
-\psi\left({1\over 2}+{1\over 2\pi T_c\tau_H}\right).
\end{equation}
We will calculate the corrections to the pair polarization bubble,
$\delta P(q,0)$, which will lead to a change in $T_c$ given by
\begin{equation}
\label{tch}
\ln{\left({T_c\over T_{c0}}\right)}=
\psi\left({1\over 2}\right)
-\psi\left({1\over 2}+{1\over 2\pi T_c\tau_H}\right)
+{\delta P(q,0)\over N(0)}
\end{equation}
Since we will assume
a featureless interaction, we need to evaluate diagrams 1 to 4 of Fig. 2.
Diagrams 1 and 3 involve the summation of three terms to form a Hikami
box\cite{Hik}, the general form for which is
\begin{equation}
\label{hikami}
2\pi N(0)\tau^4\left[-D(\De_4^1+2\De_4^2)+|\epsilon_1|+|\epsilon_2|+
|\epsilon_3|+|\epsilon_4|\right]\theta(-\epsilon_1\epsilon_2)
\theta(-\epsilon_2\epsilon_3)\theta(-\epsilon_3\epsilon_4)
\end{equation}
where $\displaystyle\De_4^1=\sum_{i=1}^4 {\bf q}_i\cdot{\bf q}_{i+1}$
and $\displaystyle\De_4^2=\sum_{i=1}^2 {\bf q}_i\cdot{\bf q}_{i+2}$,
the ${\bf q}_i$ being the incoming momenta, and the $\epsilon_i$ the
Matsubara frequencies on the electron Green functions in the box.
Using standard diagrammatic rules, and the above result for the Hikami
box, we obtain the 4 contributions to the pair polarization bubble
\begin{eqnarray}
\label{plist}
P_1&=&\displaystyle-4\pi N(0)VT\sum_{\epsilon_l}T\sum_{\epsilon_n}\sum_{q'}
{Dq^2+Dq'^2+3|\epsilon_l|+|\epsilon_n|\over (Dq^2+2|\epsilon_l|)^2
(Dq'^2+|\epsilon_l|+|\epsilon_n|)^2}\theta(-\epsilon_l\epsilon_n)
\nonumber\\
P_2&=&\displaystyle 4\pi N(0)VT\sum_{\epsilon_l}T\sum_{\epsilon_n}\sum_{q'}
{1\over (Dq^2+2|\epsilon_l|)^2(D(q'+q)^2+|\epsilon_l|+|\epsilon_n|)}
\theta(\epsilon_l\epsilon_n)\nonumber\\
P_3&=&\displaystyle -4\pi N(0)VT\sum_{\epsilon_l}T\sum_{\epsilon_n}\sum_{q'}
{Dq^2+Dq'^2+|\epsilon_l|+|\epsilon_n|\over (Dq^2+|\epsilon_l|)
(Dq^2+|\epsilon_n|)
(Dq'^2+|\epsilon_l|+|\epsilon_n|)^2}\theta(-\epsilon_l\epsilon_n)
\nonumber\\
P_4&=&\displaystyle -4\pi N(0)VT\sum_{\epsilon_l}T\sum_{\epsilon_n}\sum_{q'}
{1\over (Dq^2+2|\epsilon_l|)(Dq^2+|\epsilon_n|)
(D(q'+q)^2+|\epsilon_l|+|\epsilon_n|)}
\theta(\epsilon_l\epsilon_n)
\end{eqnarray}
We note that the relative signs of $\epsilon_l$ and $\epsilon_n$ are irrelevant
in the sum, since the summand depends only upon $|\epsilon_l|$ and
$|\epsilon_n|$,
and that the featurelessness of the potential allows us to shift the
momentum sum in terms 2 and 4. We find that terms 1 and 2 partially cancel,
whilst terms 3 and 4 reinforce, to yield the sum
\begin{equation}
\label{ptotal}
P=\displaystyle-{g\over D}T\sum_{\epsilon_l>0}T\sum_{\epsilon_n>0}
\displaystyle\left\{
{1\over [\epsilon_l+1/\tau_H][\epsilon_l+\epsilon_n]}+
{1\over[\epsilon_l+1/\tau_H][\epsilon_n+1/\tau_H]}
\left[\ln{\left({1\over[\epsilon_l+\epsilon_n]\tau}\right)}
+{1\over[\epsilon_l+\epsilon_n]\tau_H}\right]\right\}
\end{equation}
where $g=N(0)V$ and we have performed the $q'$-sum and set
$Dq^2\rightarrow 2/\tau_H$.

We first reproduce MEF's analytic approximation to the sums over
Matsubara frequencies. To do this it turns out to be easier to
make a choice of relative sign of Matsubara frequencies,
$\epsilon_l\epsilon_n<0$, and to set
$\epsilon_n=\epsilon_l+\omega_m$. If $\epsilon_l=2\pi T(l+1/2)$ and
$\omega_m=2\pi Tm$, the sum becomes
\begin{eqnarray}
\label{passum}
P&=&\displaystyle-{g\over 4\pi^2D}\sum_{m=1}^{M}\sum_{l=0}^{m-1}
\left\{{1\over [l+1/2+\alpha]m}+
{1\over[l+1/2+\alpha][m-l-1/2+\alpha]}
\left[\ln{\left({M\over m}\right)}+{\alpha\over m}\right]
\right\}\nonumber\\
&=&\displaystyle-{g\over 4\pi^2D}\sum_{m=1}^{M}
\left[\psi\left({1\over 2}+m+\alpha\right)
-\psi\left({1\over 2}+\alpha\right)\right]
\left\{{1\over m}+{2\over (m+2\alpha)}\left[\ln{\left({M\over m}\right)}
+{\alpha\over m}\right]\right\}
\end{eqnarray}
where $\alpha=1/2\pi T\tau_H$, $M=1/2\pi T\tau$. If we first evaluate
the sum for $\alpha=0$, we see that it will be dominated by large $m$,
so we replace the difference of digamma functions by $\ln{m}$, and the
sum over $m$ by an integral to get
\begin{equation}
\label{mfpapprox}
P(0)=-{g\over 4\pi^2D}\left[{1\over 3}\ln^3{\left({1\over 2\pi T_c\tau}\right)}
+{1\over 2}\ln^2{\left({1\over 2\pi T_c\tau}\right)}\right].
\end{equation}
To evaluate the result for finite $\alpha$ we subtract off the $\alpha=0$
result and try to analytically approximate the difference.
Again we expect the sum to be dominated by large $m$, so we can
ignore the $\alpha$ in $\psi(1/2+m+\alpha)$ and $(m+2\alpha)$, and ignore
the term proportional to $\alpha$ as it is less divergent. The difference
then has the form
\begin{eqnarray}
\label{mefpapprox}
P(\alpha)-P(0)&=&\displaystyle-{g\over 4\pi^2D}\sum_{m=1}^{M}
\left[\psi\left({1\over 2}\right)-\psi\left({1\over 2}+\alpha\right)\right]
\left\{{1\over m}+
{2\over m}\ln{\left({M\over m}\right)}\right\}\nonumber\\
&=&\displaystyle-{g\over 4\pi^2D}
\left[\psi\left({1\over 2}\right)-\psi\left({1\over 2}+\alpha\right)\right]
\left\{\ln{\left({1\over 2\pi T_c\tau}\right)}+
\ln^2{\left({1\over 2\pi T_c\tau}\right)}\right\}
\end{eqnarray}
From our knowledge of the calculation which includes the full screened
Coulomb interaction we know that we should set $g=1/2$, from which it
follows that $g/4\pi^2N(0)D=t$.
Putting the Eqns.~(\ref{mefpapprox}) and (\ref{mfpapprox}) into 
Eqn.~(\ref{tch}) yields the MEF formula cited in Eqn.~(\ref{mef}).
It turns out that this approximation, although apparently reasonable,
is not justified. If we plot the values of $P(\alpha)$ calculated by
performing the sums directly, we find that they do not agree with
Eqn.~(\ref{mef}). Consequently the $H_{c2}(T)$ curves predicted by the
analytic approximation and exact sum are also different.
\begin{figure}
\centerline{\psfig{figure=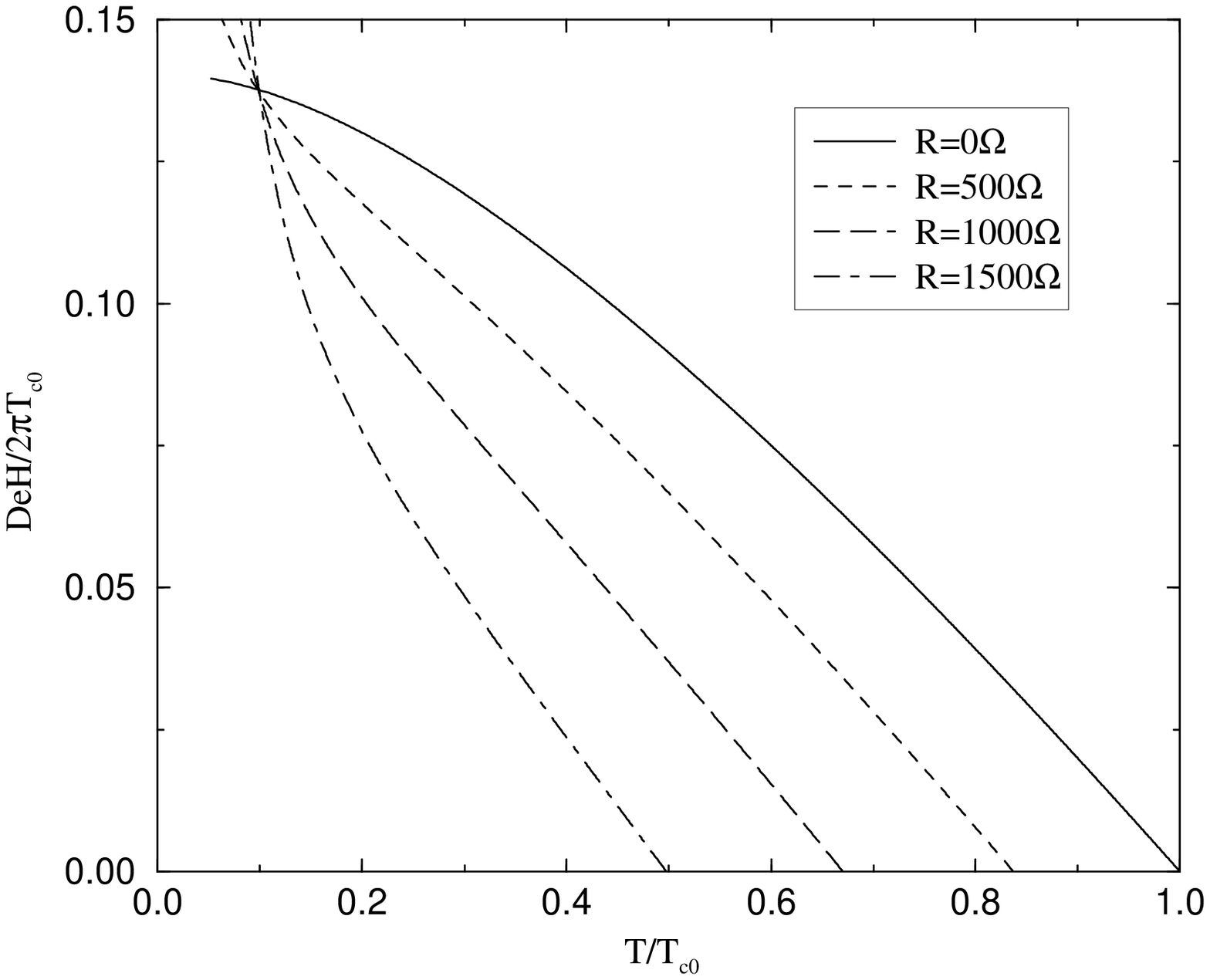,width=8.5cm}\hskip 0.25truein
\psfig{figure=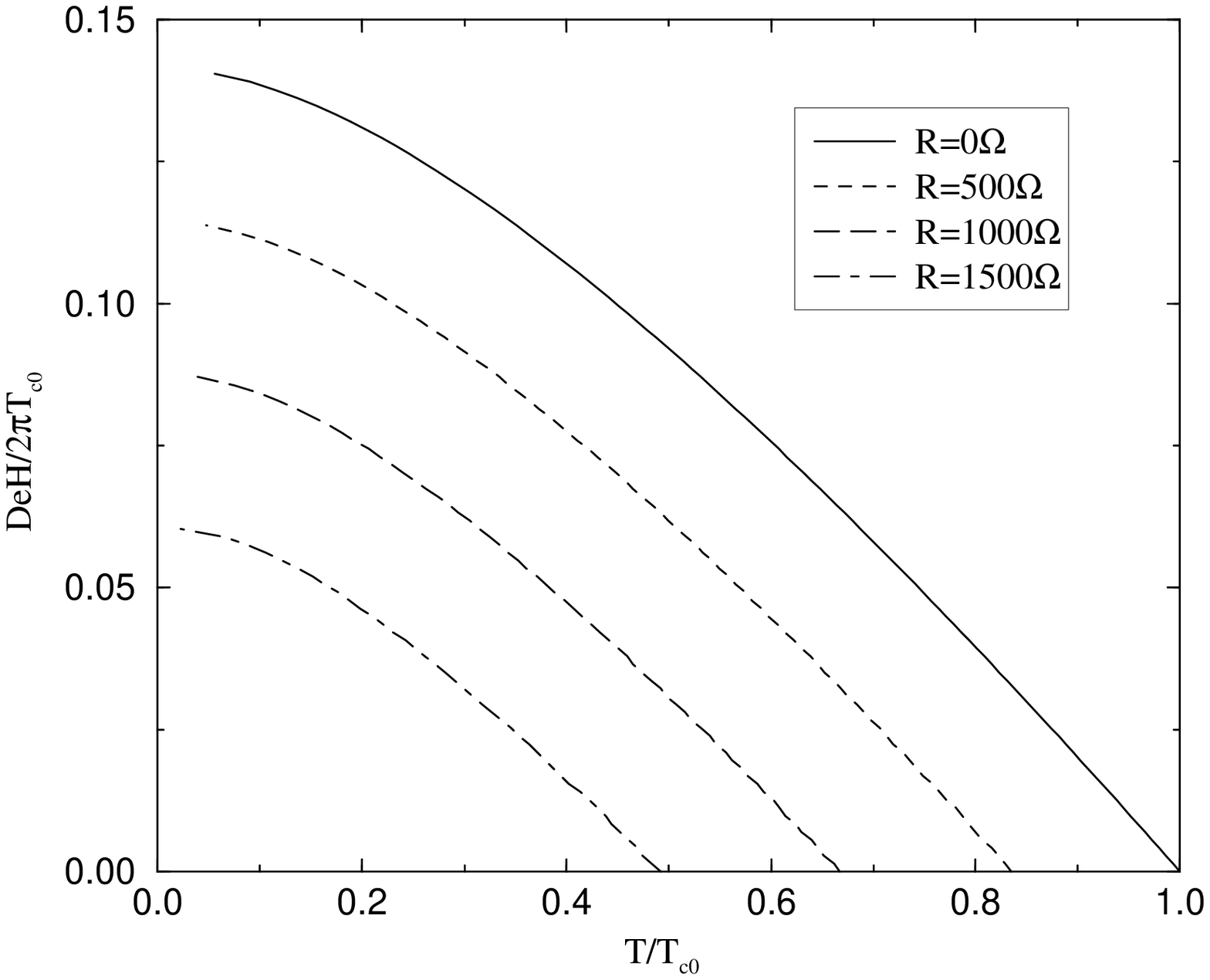,width=8.5cm}}
\caption{Plots of $H_{c2}(T)$ deduced from first-order perturbation theory
with ``self-consistency''. The curves on the left are derived using the 
analytic MEF formula, and clearly show upward curvature at low temperature
as resistance is increased. The curves on the right are derived by 
calculating Matsubara sums exactly, and show no upward curvature. It follows
that the upward curvature is an artefact of making an incorrect analytical
approximation in the derivation of the MEF formula. Both curves terminate
at non-zero values of temperature because of re-entrance problems.}
\end{figure}
\par
On the left side of Fig.~3 we display $H_{c2}(T)$ predicted by the 
MEF result of Eqn.~(\ref{mef}) for several
values of $R_{\square}$ in the case where $\ln(1/2\pi T_{c0}\tau)=6$.
We clearly see the positive curvature at low temperature, and the result 
that the curves all terminate at finite $T$ due to re-entrance problems.
On the right side of Fig.~3 we plot $H_{c2}(T)$ deduced directly from the 
first-order
perturbation theory result of Eqn.~(\ref{passum}). We note
that since the answers we get depend upon the logarithm of the upper
cut-off, if we treat the upper cut-off differently, we will get slightly
different answers. However, this difference will always be lost in fitting
to experiment since we determine the unknown upper cut-off parameter from
the initial slope of the data. The reason we mention this is that the
exact sums can be performed in two slightly different ways: we can have
the sum over $\epsilon_l$ and $\omega_m$, or the sum over
$\epsilon_l$ and $\epsilon_n$, the cut-offs in each case being taken at
$1/\tau$.
The results are slightly different due to different treatment of the
upper cut-off. However in
both cases we find no positive curvature in $H_{c2}(T)$. There is still
the problem that the curves finish at finite $T$ because of re-entrance,
and this tells us that the ad hoc prescription being used to go beyond
first-order perturbation theory is still inadequate. But again we stress
the key point of this section: even within the ad hoc extension of
first-order perturbation theory, we do not get positive curvature in
$H_{c2}(T)$ if we do the Matsubara sums exactly.

\section{Oreg and Finkelstein's Resummation Technique}

In this section we answer the question of what is the correct way
to go beyond first-order perturbation theory. To do this, we extend a
resummation technique recently developed by Oreg and Finkelstein\cite{OF} 
(OF).
This involves calculating the pair scattering amplitude,
$\Gamma_c(\epsilon_n,\epsilon_l)$, and identifying $T_c$ as the
temperature at which it first diverges. The ladder summation involved
is shown diagrammatically in Fig.~4 This is actually an extended version
of the OF approach because as well as considering diagrams
3 and 4 of Fig.~2, which correspond to an effective Coulomb pseudopotential,
and end up in the block $t\Lambda$, we also consider diagrams 1 and 2 of
Fig.~2, which correspond to an effective self-energy, and end up in the
block $\Sigma$ which renormalises the Cooperon impurity ladder. In fact we
will consider both versions of the summation technique: the simpler one
involves using the bare Cooperon $C_0$ rather than the dressed one $C$.
OF demonstrate that the matrix equation they obtain in
2D can be approximated by a parquet-like differential equation that turns
out to be identical to that obtained from renormalization group analysis.
Instead of generating a continuous approximation to the matrix equation,
we simply solve the equation and change the temperature (which affects
the upper cut-off $M=1/2\pi T\tau$) until we find the first zero eigenvalue
of the matrix, at which point it has become singular, and we have reached
$T_c$. Numerically this
involves diagonalizing matrices of rank less than 2000 or so.
\begin{figure}
\centerline{\psfig{figure=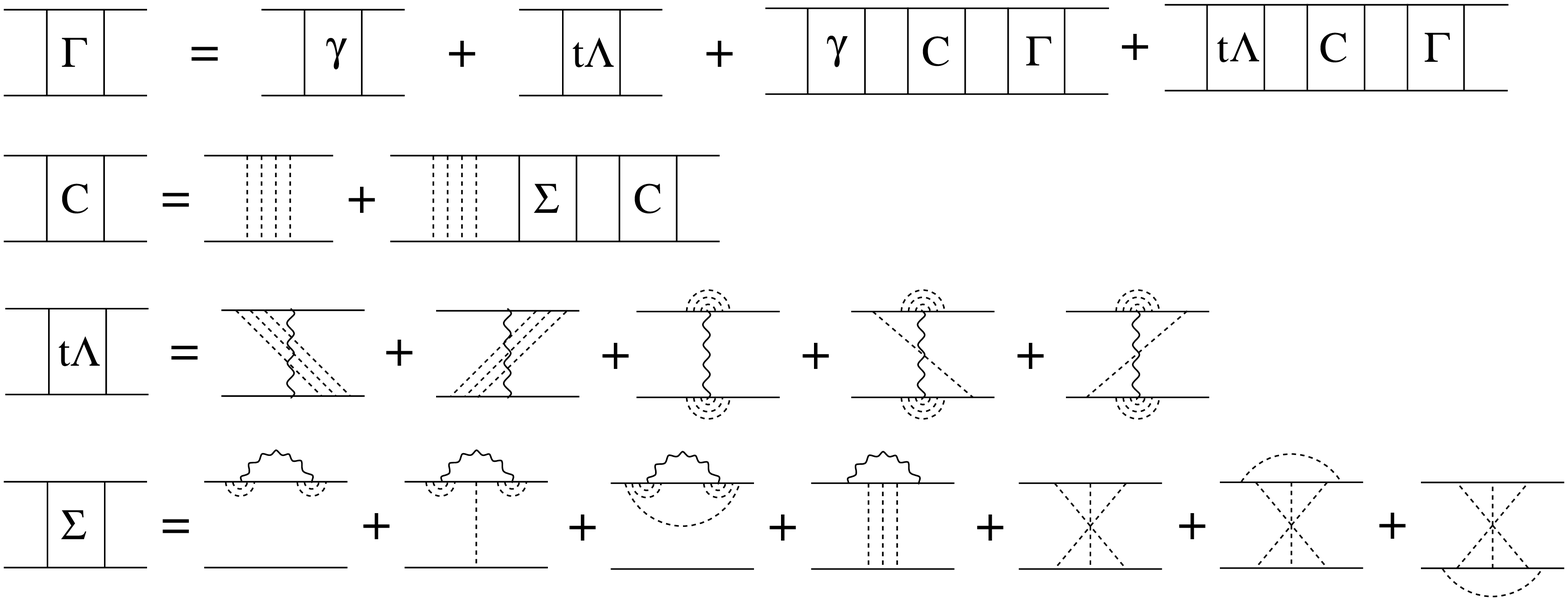,width=18cm}}
\medskip
\caption{Diagrammatic equation for the scattering amplitude matrix
$\Gamma(\epsilon_n,\epsilon_l)$. Block $\gamma$ is the BCS interaction.
Block $t\Lambda$ is the correction to the effective interaction caused by the
interplay of Coulomb interaction and disorder. Block $\Sigma$ is the 
correction to the bare Cooperon ladder $C_0$ caused by Coulomb interaction
and disorder. We consider both the original resummation procedure which
uses the bare Cooperon $C_0$, and an extended procedure which includes the
fully dressed Cooperon $C$.}
\end{figure}
\par
Let us now proceed to the calculation of the pair amplitude matrix $\Gamma$.
If we evaluate the diagrams of Fig.~4, but ignore the magnetic field and
self-energy correction to the Cooperon, we obtain the same equation as OF,
\begin{equation}
\label{OF}
\Gamma(\epsilon_n,\epsilon_l)=-|\gamma|+t\Lambda(\epsilon_n,\epsilon_l)
-\pi T\sum_{m=-(M+1)}^{M}
\left[-|\gamma|+t\Lambda(\epsilon_n,\epsilon_m)\right]
{1\over |\epsilon_m|}\Gamma(\epsilon_m,\epsilon_l)
\end{equation}
except that we have explicitly kept both positive and negative Matsubara
frequencies because the expression for $\Lambda(\epsilon_l,\epsilon_n)$
will turn out to depend upon whether
its two Matsubara frequencies have the same or opposite sign. This
is a reflection of the breaking of time-reversal invariance caused by
the magnetic field. Let us first see how to solve the above equation, and then
later put in the magnetic field, and the correction to the Cooperon.
As a matrix equation, it takes the form
\begin{equation}
{\hat\Gamma}=-|\gamma|{\hat 1}+t{\hat\Lambda}
-{1\over 2}[-|\gamma|{\hat 1}+t{\hat\Lambda}]{\hat\epsilon}^{-1}{\hat\Gamma}
\end{equation}
where
${\hat\Gamma}_{nm}=\Gamma(\epsilon_n,\epsilon_m)$, ${\hat 1}_{nm}=1$,
${\hat\Lambda}_{nm}=\Lambda(\epsilon_n,\epsilon_m)$ and
${\hat\epsilon}_{nm}=(n+1/2)\delta_{nm}$. This has the solution
\begin{equation}
{\hat\Gamma}={\hat\epsilon}^{1/2}({\hat I}-|\gamma|{\hat\Pi})^{-1}
{\hat\epsilon}^{-1/2}(-|\gamma|{\hat 1}+t{\hat\Lambda})
\end{equation}
where
\begin{equation}
{\hat\Pi}={1\over 2}{\hat\epsilon}^{-1/2}
[{\hat 1}-|\gamma|^{-1}t{\hat\Lambda}]{\hat\epsilon}^{-1/2}
\end{equation}
and ${\hat I}_{nm}=\delta_{nm}$ is the identity matrix. It follows that
when the matrix ${\hat\Pi}$ has an eigenvalue equal to $1/|\gamma|$, the
pair amplitude diverges, and we have found $T_c$. Our approach is therefore
to start at the mean field transition temperature, $T_{c0}$, and decrease
temperature until one of the eigenvalues of $({\hat I}-|\gamma|{\hat\Pi})$
changes sign. The matrix ${\hat\Pi}(T)$ depends upon $T$ both through the
dependence of $\hat\Lambda$ upon $T$ and through its  rank $2M$, where
$M=(2\pi T\tau)^{-1}$. We start at the mean-field value of $M$, which we will
call $M_0$, and decrease
the temperature by increasing $M$ successively by one. 
We diagonalize the matrix $\hat\Pi$
for each value of $M$ until an eigenvalue changes sign. At this point we have
found $T_c$ for the given problem, and $T_c/T_{c0}=M_0/M$. We can then change
a parameter such as $t$ or $\alpha$ and repeat the procedure. This removes
the need for any perturbative expansion.
\par
Let us now put back in the magnetic field and Cooperon self-energy corrections.
The $|\epsilon_m|$ denominator in the Eqn.~(\ref{OF}) comes from the Cooperon
\begin{equation}
C_0(\epsilon_m)={1\over 2\pi N(0)\tau^2}
{1\over Dq^2+2|\epsilon_m|}
\end{equation}
and in the presence of a magnetic field, $|\epsilon_m|$ is replaced by
$|\epsilon_m|+1/\tau_H$, and hence
${\hat\epsilon}_{nm}=(n+1/2+\alpha)\delta_{nm}$, where
$\alpha=(2\pi T\tau_H)^{-1}$. The contributions
to $t\Lambda$ from the diagrams of Fig.~4 which correspond to diagrams 3 and
4 of Fig.~2 are given by
\begin{eqnarray}
t\Lambda_3(\epsilon_n,\epsilon_m)&=&\displaystyle
{g\over\pi N(0)}\sum_{q'}
{Dq^2+Dq'^2+|\epsilon_n|+|\epsilon_m|\over
[Dq'^2+|\epsilon_n|+|\epsilon_m|]}\theta(-\epsilon_n\epsilon_m)\nonumber\\
t\Lambda_4(\epsilon_n,\epsilon_m)&=&\displaystyle
{g\over\pi N(0)}\sum_{q'}{1\over [Dq'^2+|\epsilon_n|+|\epsilon_m|]}
\theta(\epsilon_n\epsilon_m)
\end{eqnarray}
and so performing the $q'$-sum and setting $Dq^2\rightarrow 2/\tau_H$ gives
\begin{equation}
t\Lambda(\epsilon_n,\epsilon_m)={g\over 4\pi^2N(0)D}
\cases{\displaystyle\ln{\left[{1\over (|\epsilon_n|+|\epsilon_m|)\tau}\right]}
&\quad$\epsilon_n\epsilon_m>0$\cr
\displaystyle\ln{\left[{1\over (|\epsilon_n|+|\epsilon_m|)\tau}\right]}
+{2\over (|\epsilon_n|+|\epsilon_m|)\tau_H}&\quad
$\epsilon_n\epsilon_m<0$\cr}
\end{equation}
It follows that the matrix elements of $\hat\Lambda$ are
\begin{equation}
\Lambda_{nm}=\cases{
\displaystyle\ln{\left({M\over n+m+1}\right)}&\quad$\epsilon_n\epsilon_m>0$\cr
\displaystyle\ln{\left({M\over n+m+1}\right)}+{2\alpha\over (n+m+1)}&
\quad$\epsilon_n\epsilon_m<0$\cr}
\end{equation}
\par
To include the self-energy correction into the Cooperon, we note
that the corrected Cooperon is given by
\begin{equation}
C=[C_0^{-1}-\Sigma]^{-1}=
{1\over 2\pi N(0)\tau^2}\displaystyle
\left[Dq^2+2|\epsilon_m|-{1\over 2\pi N(0)\tau^2}\Sigma\right]^{-1}
\end{equation}
so that we need to absorb a factor $-1/2\pi N(0)\tau^2$ into $\Sigma$.
The contributions from the diagrams of Fig.~4 which correspond to
diagrams 1 and 2 of Fig.~2 are
\begin{eqnarray}
\Sigma_1(\epsilon_n)&=&\displaystyle{2g\over N(0)}T\sum_{\epsilon_m}\sum_{q'}
\left[{Dq^2+Dq'^2+3|\epsilon_n|+|\epsilon_m|\over
(Dq'^2+|\epsilon_n|+|\epsilon_m|)^2}\right]\theta(-\epsilon_n\epsilon_m)
\nonumber\\
\Sigma_2(\epsilon_n)&=&\displaystyle-{2g\over N(0)}T\sum_{\epsilon_m}\sum_{q'}
{1\over (Dq'^2+|\epsilon_n|+|\epsilon_m|)}\theta(\epsilon_n\epsilon_m)
\end{eqnarray}
which partially cancel to give the result
\begin{equation}
\Sigma(\epsilon_n)=\displaystyle{g\over 2\pi N(0)D}T\sum_{m}
{1\over (|\epsilon_n|+|\epsilon_m|)}\theta(-\epsilon_n\epsilon_m)
=\displaystyle{g\over 4\pi^2 N(0)D}\left(\sum_{k=n+1}^{M} {1\over k}\right)
[Dq^2+2|\epsilon_n|]
\end{equation}
The weak-localization contribution to the Cooperon self-energy is given by
\begin{equation}
\Sigma_{WL}(\epsilon_n)\displaystyle=-{1\over 2\pi N(0)}\sum_{q'}
{2D(q^2+q'^2)+4|\epsilon_n|\over Dq'^2+2|\epsilon_n|}
=\displaystyle-{1\over 8\pi^2N(0)D}
\ln{\left({1\over 2|\epsilon_n|\tau}\right)}[Dq^2]
\end{equation}
Incorporating $\Sigma$ into the Cooperon means that $\hat\epsilon$ has
elements
\begin{equation}
{\hat\epsilon}_{nm}=\left\{\left(n+{1\over 2}+\alpha\right)
\left[1+t\sum_{k=n+1}^{M} {1\over k}\right]
-2\alpha t\ln{\left({M\over n+1/2}\right)}\right\}\delta_{nm}
\end{equation}
Using the new formulas for $\hat\epsilon$ and $\hat\Lambda$, we now plot
$H_{c2}(T)$ for various values of $t=R_{\square}/R_0$. The results are shown in
Fig.~5: the plot on the left does not include self-energy corrections;
the plot on the right does. The
results are very similar and show no sign of upward curvature in $H_{c2}(T)$.
We note that there are no re-entrance problems in our calculation of $T_c$:
we can plot the curves down to as low a temperature as we like if we are
prepared to diagonalize large enough matrices.
\begin{figure}
\centerline{\psfig{figure=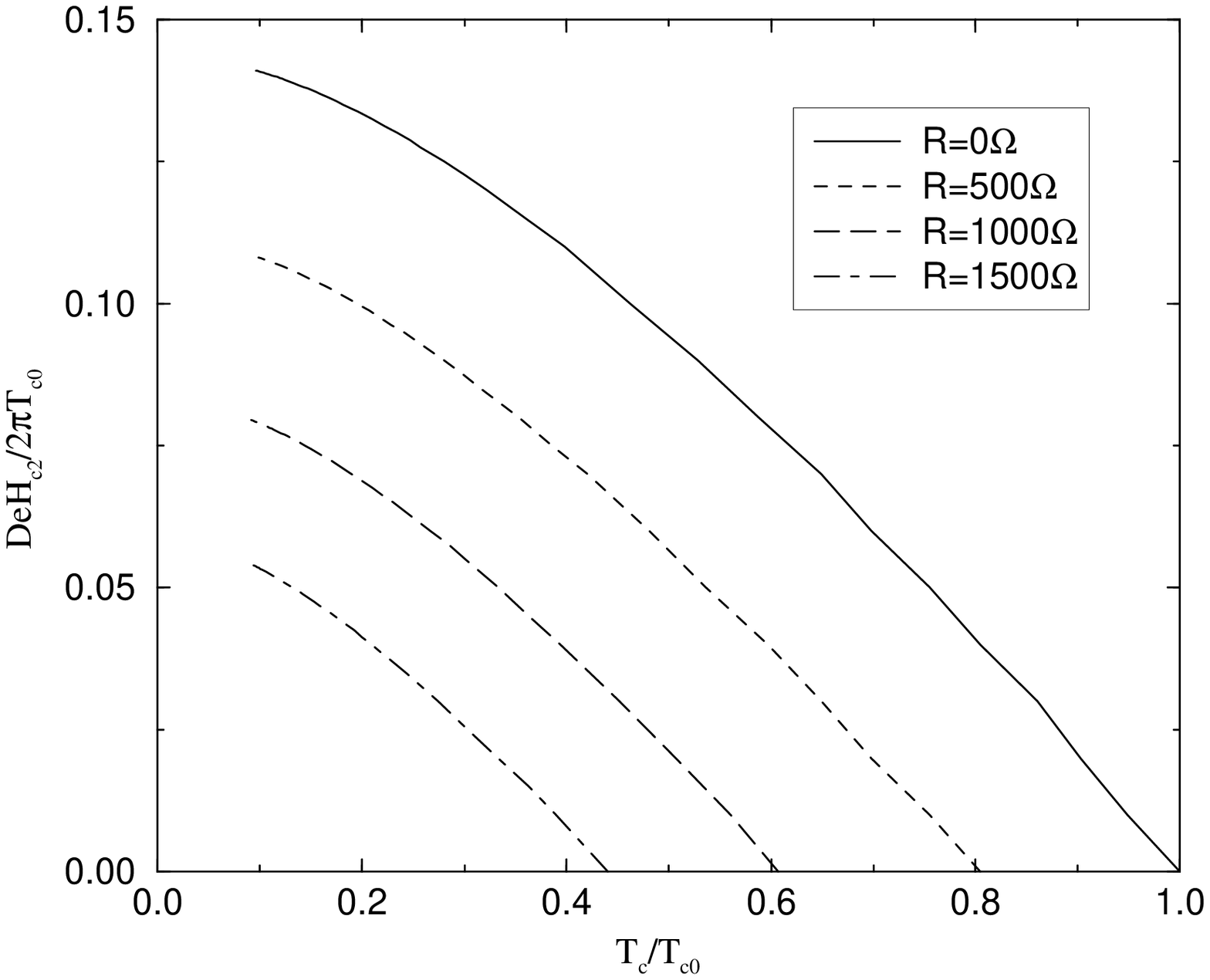,width=8.5cm}\hskip 0.25truein
\psfig{figure=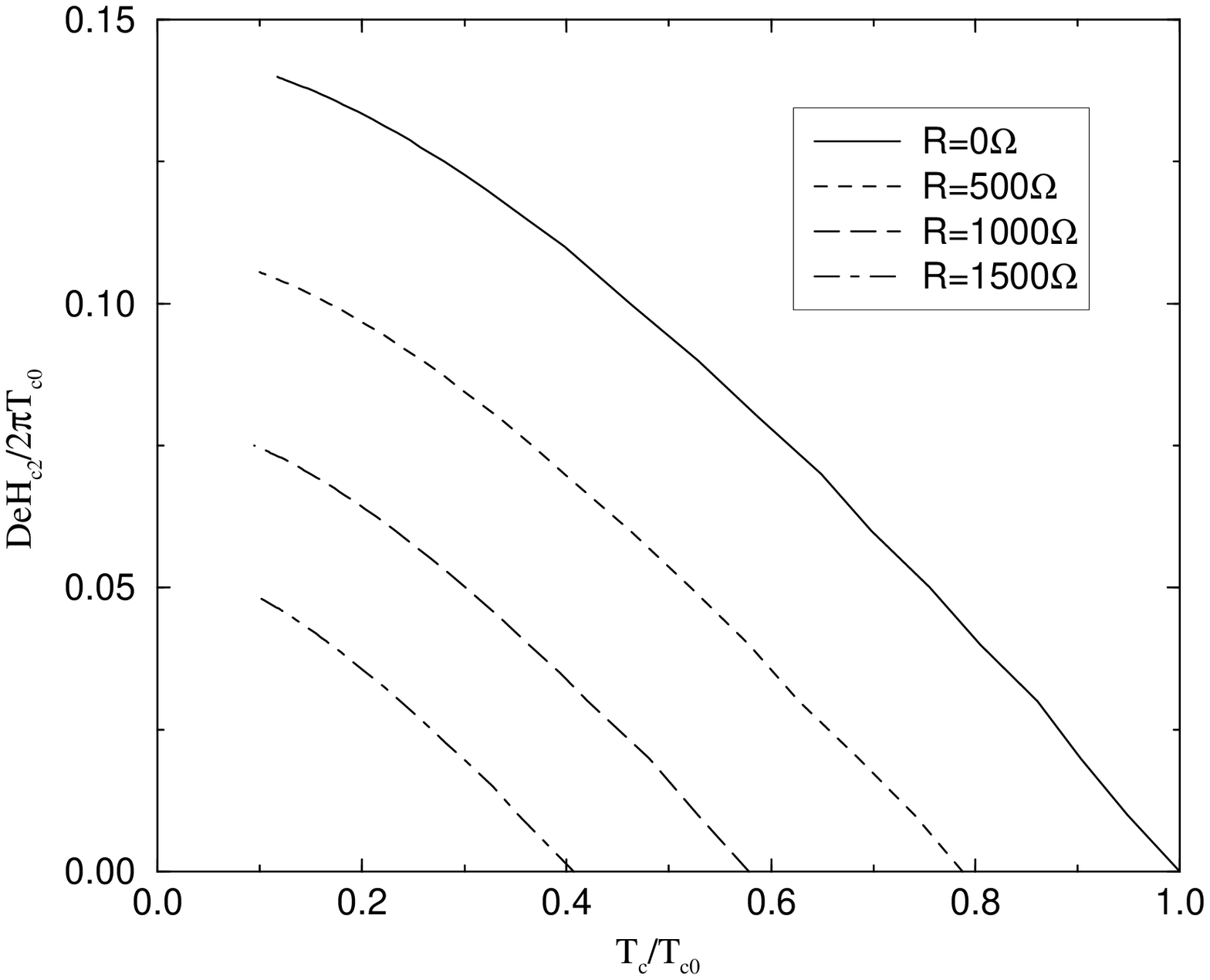,width=8.5cm}}
\caption{Plots of $H_{c2}(T)$ deduced from the non-perturbative
resummation method. The curve on the right includes the self-energy
corrections to the Cooperon ladder; the curve on the left does not.
Neither curve show any sign of positive curvature at low temperature.}
\end{figure}

\section{Discussion and Conclusions}

The central message of the paper is that the theory of localization and
interaction does not predict an anomalous positive curvature in the upper
critical field $H_{c2}(T)$ of thin film superconductors. A subsidiary
message is that the resummation method developed by 
Oreg and Finkelstein\cite{OF}
is a very powerful and adaptable tool for going beyond perturbation theory
in disordered superconductors. We suspected that the positive curvature in
$H_{c2}(T)$ was an artefact related to the re-entrance problem found in
$T_c(R_{\square})$, and that it would not survive a systematic
non-perturbative treatment. The latter is true, but the cause of the
positive curvature artefact turns out to be an incorrect analytic
approximation to sums of Matsubara frequencies.
\par
We will now consider the experimental situation by attempting to fit
experimental data to both the MEF formula of Eqn.~(\ref{mef}), and the 
exact first-order perturbation theory result of Eqn.~(\ref{passum}).
The data which fits best to the
localization theory is that of Graybeal and Beasley\cite{GB} on amorphous
films of Mo-Ge, because values
of $T_c$ for all films can be obtained by choosing a single value of the
parameter $\beta=\ln{(1/2\pi T_{c0}\tau)}$. The fits in Fig.~6 used the 
parameter value $\beta=7.17$ for the MEF curves and $\beta=5.3$ for the exact
sum curves. Values of $T_{c0}=7.2K$ and $H_{c0}=120$kOe were taken directly
from the data. The experimental data appears to fit better to the exact sum
curves than the MEF curves, and there seems little sign of an upward curvature
in $H_{c2}(T)$. 
\par
When the same procedure is applied to the data of Okuma et al\cite{OKOK}
on Zn films, we find that we need to use different values of $\beta$ for
the two films to get the correct $T_c$ for given $R_{\square}$. The plots
in Fig.~7 use values
$\beta=7.5$ for the $400\Omega$ film and $\beta=8.1$ for the $600\Omega$ film
for the MEF formula; $\beta=5.56$ for the $400\Omega$ film and 
$\beta=6.1$ for the $600\Omega$ film for the exact sum formula. 
Both films have the same thickness
of $100\AA$, but different resistances, and hence different diffusion constants
$D$. The latter were calculated using the material parameters for Zn to get
values of $H_{c2}(T=0)$ for zero-resistance films of the same composition
equal to 5.08kOe for the $400\Omega$ film and 7.62kOe for $600\Omega$ film. 
$T_{c0}$
was taken to be $1K$. Again the data seems to fit better to the exact sum
curves than the MEF formula.
\par
To fit the data of Hebard and Paalanen\cite{HP} on amorphous
In-InO$_x$ films again requires a different value of $\beta$ for each film. The plots
in Fig.~8 use
$\beta=3.5$ for the $2250\Omega$ film and $\beta=4.0$ for the $2900\Omega$ and
$3300\Omega$ films for the MEF curves; $\beta=1.6$ for $2250\Omega$ film and
$\beta=2.0$ for the $2900\Omega$ and $3300\Omega$ films for the exact sum
curves. The value of $T_{c0}=3.6K$ quoted in the paper was used, and a value 
of $H_{c2}(0)=100kOe$ was estimated from the experimental plots.
There is a clear upturn in $H_{c2}(T)$ at low $T$, but it is not explained
by the MEF curves which are effectively all upturn at these parameter values.
Again the fit to exact sum curves seems better.
\par
In conclusion it seems that the experimental curves fit better to the
exact sum curves than the MEF curves. Any low temperature upturns in 
$H_{c2}(T)$ are not explained by the MEF formula, and presumably correspond
to different physics.

\begin{figure}
\centerline{\psfig{figure=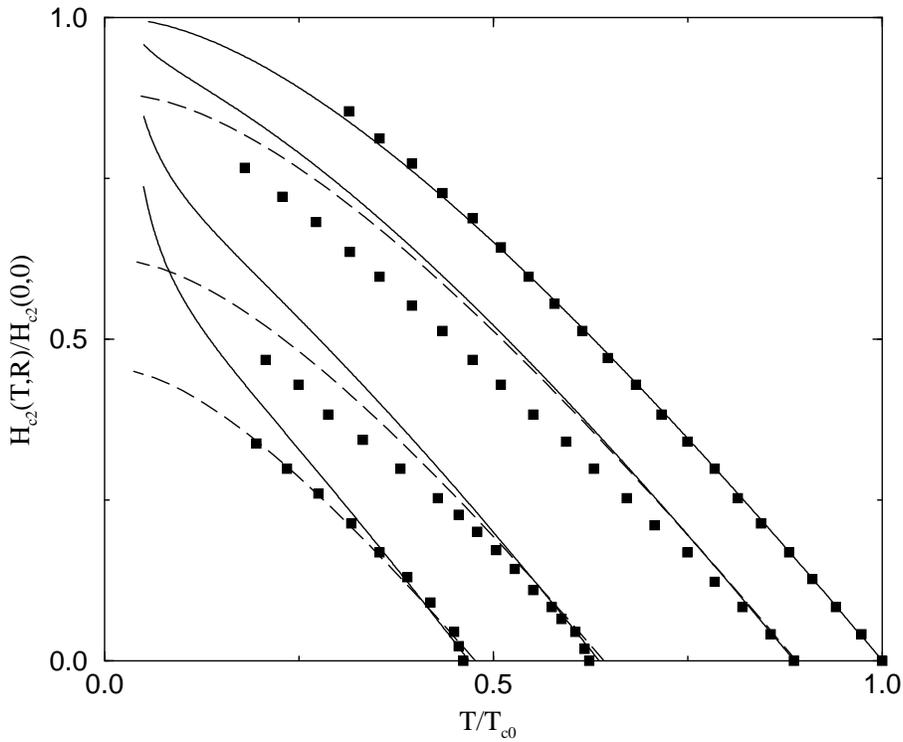,width=12cm}}
\caption{Fitting of Graybeal and Beasley's data\cite{GB} to the MEF
formula (solid lines) and perturbation theory with exact sums 
(dashed lines). In each case only a single fitting parameter is needed
for all curves, corresponding to the initial slope of the $T_c(R_{\square})$
curve. The films have resistances $R_{\square}\approx 0,110,400,600\Omega$.}
\end{figure}

\begin{figure}
\centerline{\psfig{figure=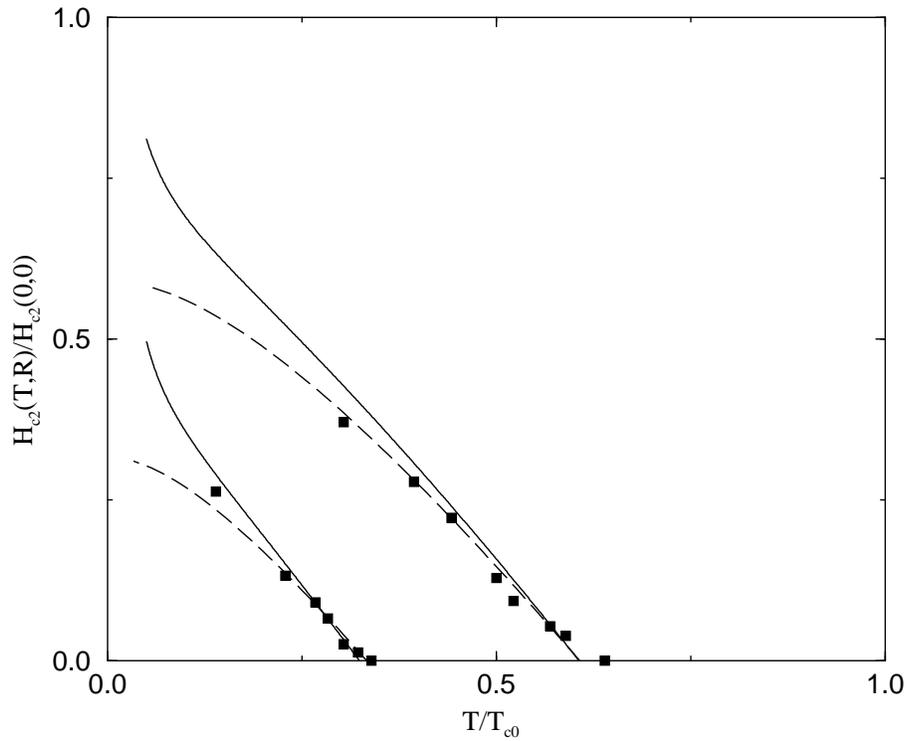,width=12cm}}
\caption{Fitting of Okuma et al's data\cite{OKOK} to the MEF
formula (solid lines) and perturbation theory with exact sums
(dashed lines). Different values of the fitting parameter are needed
for the two films to get the correct value of $T_c$ for the given
$R_{\square}$. The films have resistances $R_{\square}=400,600\Omega$.} 
\end{figure}

\begin{figure}
\centerline{\psfig{figure=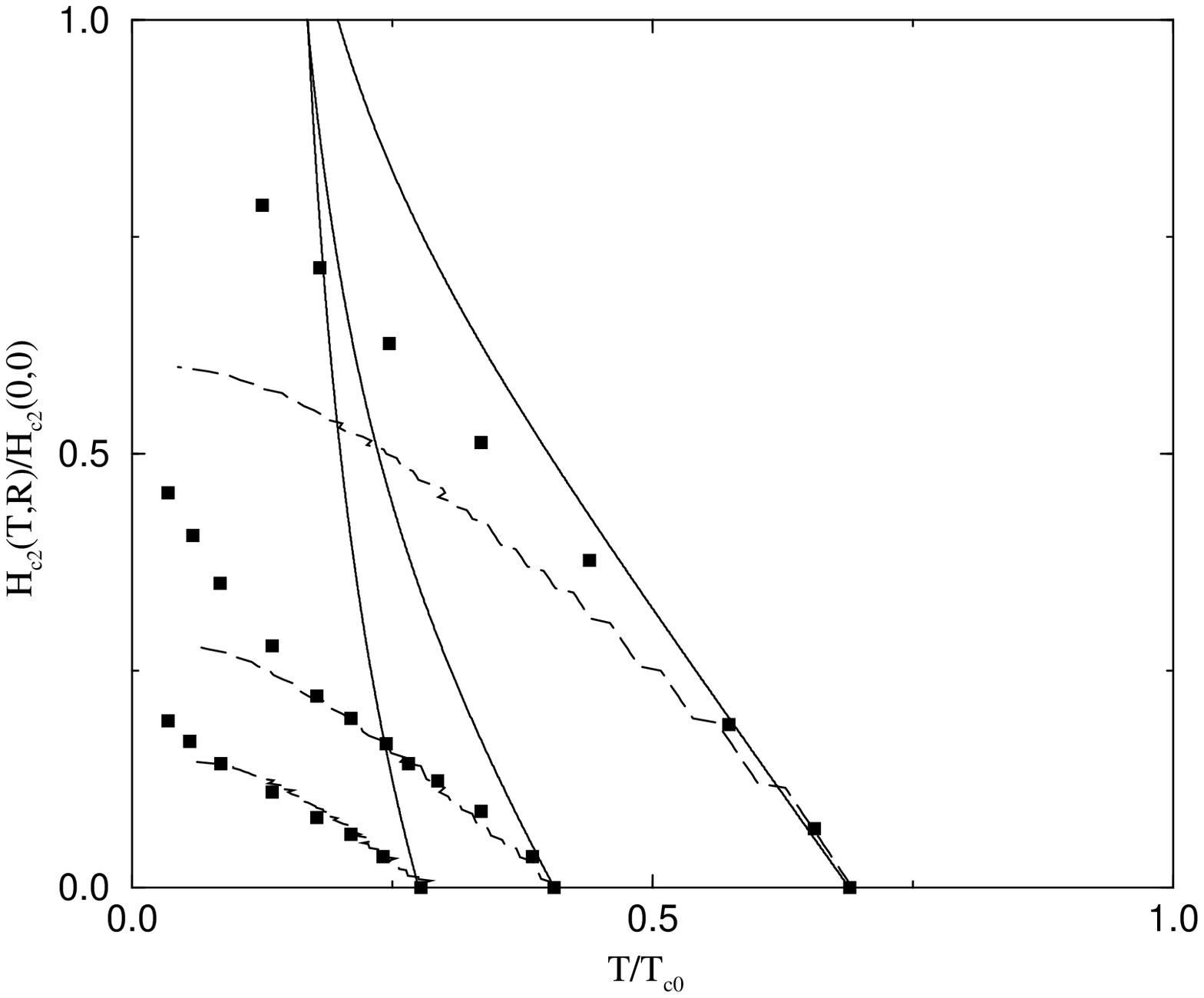,width=12cm}}
\caption{Fitting of Hebard and Paalanen's data\cite{HP} to the MEF
formula (solid lines) and perturbation theory with exact sums
(dashed lines). Different values of the fitting parameter are needed
for the three films to get the correct value of $T_c$ for the given
$R_{\square}$. The films have resistances $R_{\square}=2250,2900,3300\Omega$.}
\end{figure}

\bigskip
\centerline {\bf ACKNOWLEDGEMENTS}
\medskip

We thank I. Aleiner, A. Clerk, A.M. Finkel'stein, I.V. Lerner and Y. Oreg
for helpful discussions.
RAS acknowledges the support of a Nuffield Foundation Award to
Newly Appointed Lecturers in Science and Mathematics.
BSH acknowledges the support of a UK EPSRC Graduate Studentship.
VA is supported by the US National Science Foundation under grant
DMR-9805613.

\end{document}